\documentclass[%
 preprint,
 amsmath,amssymb,
 aps,
prl,
]{revtex4-2}
\usepackage{graphicx}
\usepackage{dcolumn}
\usepackage{bm}
\usepackage{braket}

\usepackage[separate-uncertainty = true,
  multi-part-units = repeat]{siunitx}

\begin{document}

\section{Title}
\noindent
Probing temperature dependent non-radiative decay channels in CrSBr with nonlinear optics

\section{Author list}
\noindent
Minjiang Dan$^{1,2,3,10}$, Till Weickhardt$^{2,10}$, Paul Herrmann$^2$,  Fabian Glatz$^2$,  Marie-Christin Heißenbüttel$^4$, Thorsten Deilmann$^4$, Michael Rohlfing$^4$, Ksenia Mosina$^5$, Zden\v ek Sofer$^5$, Benjamin Pingault$^{6,7}$, Julian Klein$^{8,\star}$, Giancarlo Soavi$^{2,9,\dagger}$

\section{Affiliations}
\noindent
$^1$Joint Laboratory for Extreme Conditions Matter Properties, School of Mathematics and Physics, Southwest University of Science and Technology, Mianyang 621010, China
\newline
$^2$Institute of Solid State Physics, Friedrich Schiller University Jena, Helmholtzweg 5, 07743 Jena, Germany
\newline
$^3$School of Physics, University of Electronic Science and Technology of China, Chengdu 610054, China
\newline
$^4$Institute of solid state theory, University of Münster, 48149 Münster, Germany
\newline
$^5$Department of Inorganic Chemistry, University of Chemistry and Technology Prague, Technickà 5, 166 28, Prague 6, Czech Republic
\newline
$^6$Materials Science Division and Q-NEXT, Argonne National Laboratory, Lemont, IL, USA
\newline
$^7$Chicago Quantum Institute and Pritzker School of Molecular Engineering, University of Chicago, Chicago, IL, USA
\newline
$^8$Department of Materials Science and Engineering, Massachusetts Institute of Technology, Cambridge, Massachusetts 02139, USA
\newline
$^{9}$Abbe Center of Photonics, Friedrich Schiller University Jena, Albert-Einstein-Straße 6, 07745 Jena, Germany
\newline
$^{10}$These authors contributed equally to this work. 
\newline
$^{\star}$ jpklein@mit.edu
$^{\dagger}$ giancarlo.soavi@uni-jena.de


\section{Abstract}
Among the recently discovered layered magnetic materials, the antiferromagnet CrSBr has attracted considerable interest thanks to its stability, quasi-one-dimensional crystal structure, and direct near-infrared band gap. These properties make it a suitable platform for combined electrical, magnetic, and optical measurements, and thus for studying exciton dynamics and transport across different magnetic phases. In this work, we investigate the optical response of CrSBr across its paramagnetic and antiferromagnetic phases by combining nonlinear optical spectroscopy, including second-harmonic generation, third-harmonic generation, and two-photon photoluminescence, with linear photoluminescence measurements. The observation of a clear difference in the temperature dependence of third-harmonic generation and photoluminescence intensities allows us to identify the onset of non-radiative recombination channels in the antiferromagnetic phase.

\section{Main text}
\subsection{\label{sec:1}Introduction}
The family of layered materials has recently expanded to include layered magnets, \textit{i.e.} materials that possess layer dependent magnetic properties down to the monolayer limit~\cite{Mak2019,kurebayashi2022magnetism}. To date, different types of layered magnetic materials have been explored, including ferromagnetic~\cite{huang2017layer} and antiferromagnetic~\cite{freitas2013antiferromagnetism,goser1990magnetic} phases, with both in-plane~\cite{bonilla2018strong} and out-of-plane \cite{o2018room} spin alignments. Among layered magnets, CrSBr has attracted great attention thanks to its chemical, optical and symmetry properties. It is an A-type antiferromagnet~\cite{telford2020layered} with a direct electronic layer dependent bandgap of 1.5 - \SI{2.0}{\electronvolt}~\cite{telford2020layered,klein2023bulk,Bianchi2023}, making it ideal for optical experiments in the visible/near-IR spectral range. Its crystalline structure displays a strong structural anisotropy, leading to a quasi-one-dimensional quantum confinement~\cite{klein2023bulk,wu2022quasi}. Furthermore, CrSBr exhibits excellent air stability and a relatively high N\'eel temperature of $\approx$\SI{132}{\kelvin} \cite{goser1990magnetic,wang2019family,torres2023probing}, allowing the study of few-layer samples and access into a magnetic phase diagram through temperature-dependent experiments. Finally, thanks to the strong exciton-phonon \cite{klein2023bulk,lin2024strong}, exciton-magnon \cite{bae2022exciton,dirnberger2023magneto}, magneto-electronic \cite{wilson2021interlayer, markina2025} and spin-phonon \cite{torres2023probing,pawbake2023raman,xu2022strong} coupling, CrSBr has also emerged as a promising platform for applications in optics, spintronics and optoelectronics. 

To further advance our understanding of the optical properties of CrSBr and their impact on possible technological applications in photonics and optoelectronics, it is important to gain insights into the interplay between absorption and radiative emission in the paramagnetic (PM) and antiferromagnetic (AFM) phases. To this end, nonlinear optics (NLO) is a powerful tool to study electronic symmetries and resonances in layered materials~\cite{dogadov2022parametric}. With transition metal dichalcogenides (TMDs), NLO is already an established probe of crystal symmetries~\cite{wen2019nonlinear}, excitonic states~\cite{wang2015giant}, number of layers~\cite{li2013probing,malard2013observation}, strain~\cite{mennel2018optical}, the valley degree of freedom~\cite{herrmann2023nonlinear, herrmann2025valley, friedrich2025}, and quantum geometry~\cite{tornow2026}. NLO is also a unique spectroscopy tool to study ferroic materials~\cite{fiebig2023nonlinear}. In the context of layered magnets, second-harmonic generation (SHG) has been used to study the AFM phase in bilayer CrI$_{\rm{3}}$~\cite{sun2019giant}, and to detect AFM domains and the magnetoelectric effect in MnPS$_{\rm{3}}$~\cite{ni2021direct,chu2020linear}. For CrSBr, NLO has been used to study the PM to AFM phase transition and magnetic order for samples with varying layer numbers~\cite{lee2021magnetic}.

Building on these seminal results, in this work we apply NLO to study the energy excited state landscape in bulk CrSBr. Specifically, we study resonant enhancement and temperature dependence of excitonic transitions, through SHG and third harmonic generation (THG) over a broad photon energy range (1.2 - \SI{2.6}{\electronvolt}). While SHG is strongly modulated by temperature (up to a factor of 25 at \SI{1.80}{\electronvolt}) at all investigated photon energies, due to the appearance of a time-noninvariant contribution to the nonlinear optical susceptibility below the N\'eel temperature~\cite{lee2021magnetic}, THG is almost independent of temperature in the range from \SI{40}{\kelvin}, \textit{i.e.}, low temperature (LT), to room temperature (RT, $\approx$ 300 K), and almost at any photon energy, except at resonance with the lowest excitonic state at \SI{1.28}{\electronvolt}. For this specific photon energy, we observe a temperature dependence for TH intensity that is opposite compared to that observed for the photoluminescence (PL) intensity. 
The TH intensity is higher at LT compared to RT, while the PL intensity increases monotonically for increasing temperature. We discuss this difference and ascribe it to the emergence of a non-radiative decay channel at low temperatures. With this, we contribute to the debate about the landscape of bright and dark excitons in CrSBr~\cite{Krelle2025}.

\subsection{\label{sec:2}Sample fabrication and optical characterization}

\begin{figure}
\includegraphics[width=0.8\textwidth]{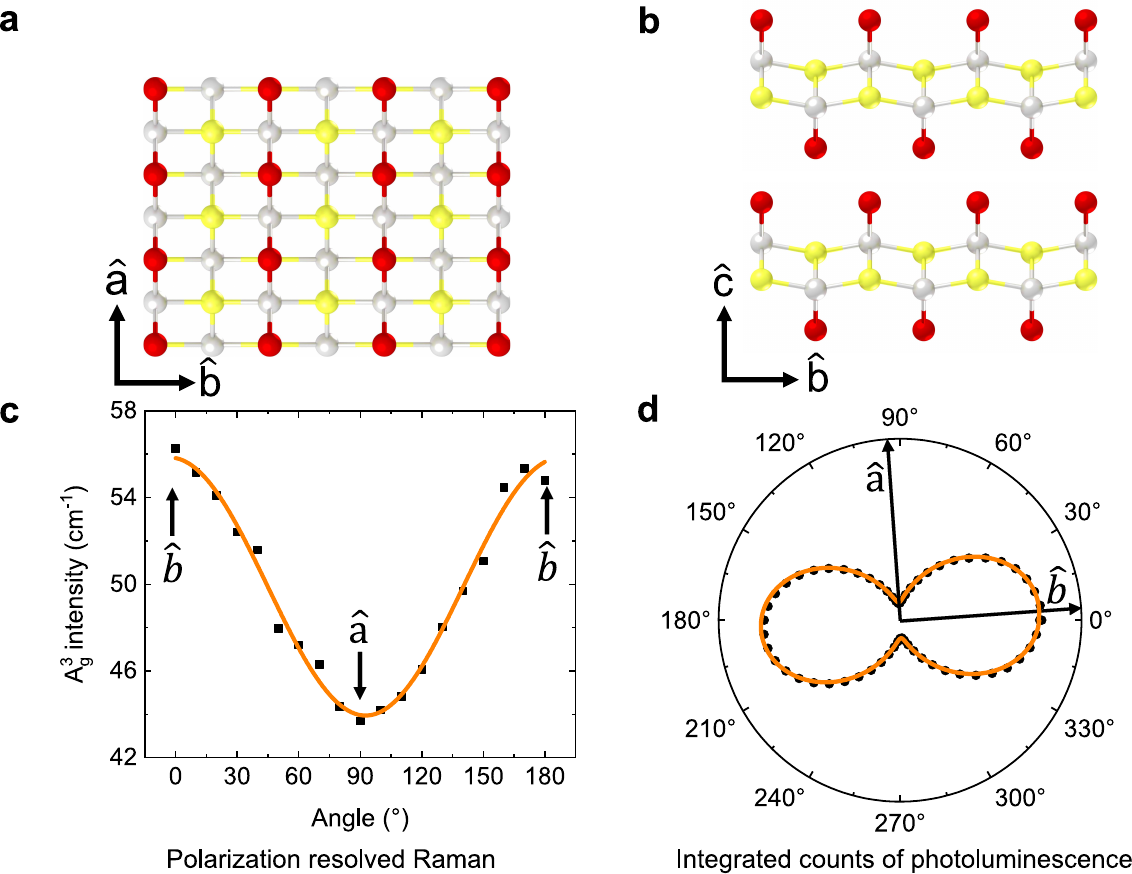}
\caption{\label{fig1}\textbf{Crystal structure of CrSBr}. Sketch of the crystal structure viewed along the $\hat{\mathrm{c}}$-axis (\textbf{a}), and the $\hat{\mathrm{a}}$-axis (\textbf{b}). Cr, S and Br atoms are depicted in gray, yellow and red spheres, respectively.
\textbf{c}, Polarization dependence of the A$_{\rm{g}}^3$ Raman mode excited with a \SI{2.33}{\electronvolt} laser showing a minimum along the $\hat{\mathrm{a}}$-axis and the maximum along the $\hat{\mathrm{b}}$-axis. \textbf{d}, Polarization dependence of the PL intensity (integrated counts in the energy range 1.28 - \SI{1.38}{\electronvolt}, excited with \SI{2.33}{\electronvolt} at room temperature) showing a clear two-fold pattern with maximum along the $\hat{\mathrm{b}}$-axis.
}
\end{figure}

\begin{figure}
\includegraphics[width=\textwidth]{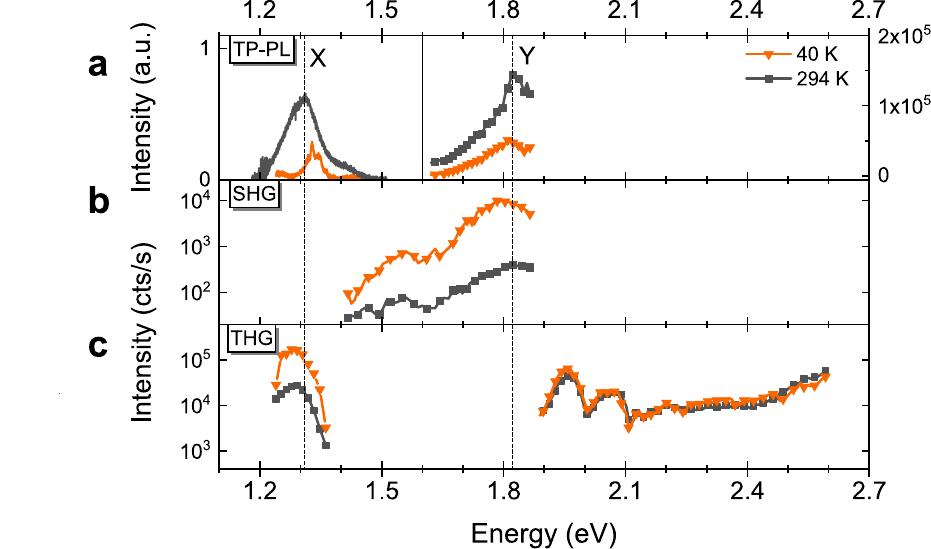}
\caption{\label{fig2}\textbf{NLO spectroscopy of bulk CrSBr}. 
Two-Photon-PL (\textbf{a}), SHG (\textbf{b}) and THG (\textbf{c}) intensities at \SI{40}{\kelvin} and \SI{300}{\kelvin} at varied emission photon energies.
The top left panel of \textbf{a} shows the Two-Photon PL spectra for an excitation photon energy of \SI{0.91}{\electronvolt}. Two excitonic features X and Y are indicated at \SI{1.31}{\electronvolt} and \SI{1.82}{\electronvolt}. The excitation polarization is not aligned with one of the in-plane crystal axes.}
\end{figure}

Bulk crystals of CrSBr were grown by chemical vapor transport (see Supplementary Information S1 and Ref.~\cite{Klein2022}) and thinned using mechanical exfoliation onto a SiO$_{\rm{2}}$(\SI{90}{\nano\meter})/Si substrate. The exfoliated CrSBr samples have thicknesses in the range 15 to \SI{70}{\nano\meter}, as measured by atomic force microscopy (Supplementary Information S2). Fig.~\ref{fig1}a-b schematically show the crystal structure of bulk CrSBr, with an orthorhombic symmetry when viewed along the $\hat{\mathrm{c}}$-axis (stacking axis through van der Waals (vdW) interaction, Fig.~\ref{fig1}a)~\cite{deng2023situ}, and Br atoms protruding to the top and bottom of the layer (Fig.~\ref{fig1}b). The orientation of the crystal axes were confirmed by polarization resolved Raman (Fig.~\ref{fig1}c) and PL (Fig.~\ref{fig1}d) measurements both of which are strongest along the $\hat{\mathrm{b}}$-axis due to the anisotropic electronic structure. 

\begin{figure}
\includegraphics[width=0.8\textwidth]{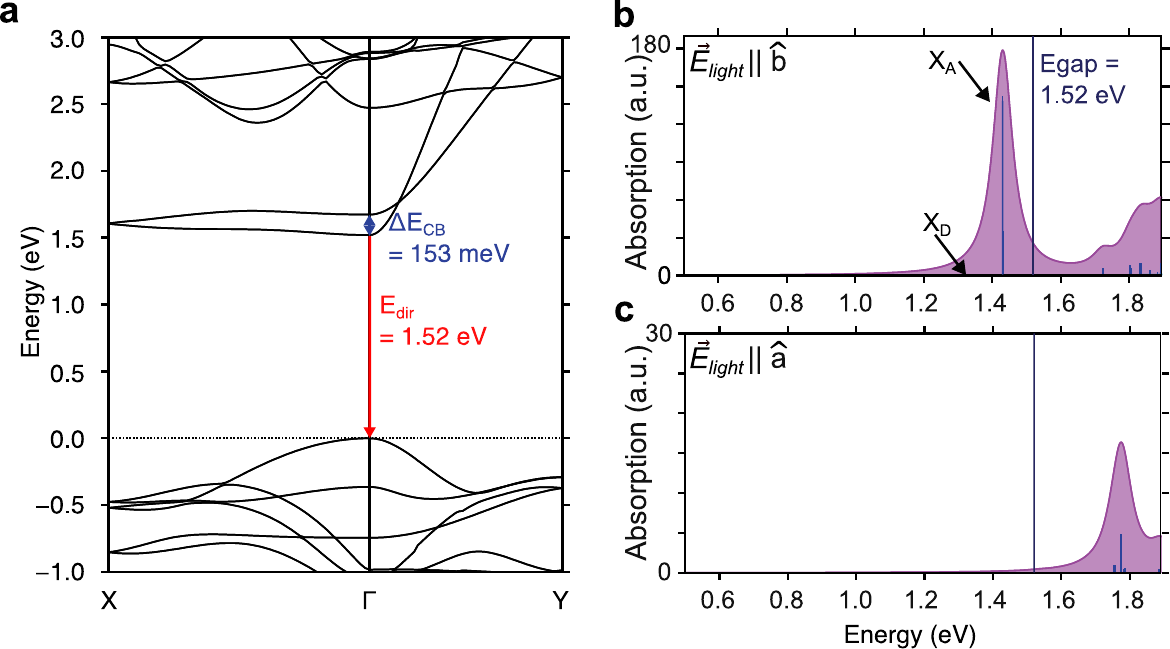}
\caption{\label{fig3}\textbf{Band structure and excitons in bulk CrSBr}. \textbf{a} Bandstructure of the AFM bulk calculated from $GW$ theory. The anisotropic lowest bands are split by \SI{153}{\milli\electronvolt} and the band gap is \SI{1.52}{\electronvolt}.
\textbf{b} and \textbf{c} show the optical absorption from the solution of the BSE for light polarized along the $\hat{\mathrm{b}}$ and the $\hat{\mathrm{a}}$ crystal axis respectively.}
\end{figure}

Fig.~\ref{fig2} shows the TP-PL, SHG and THG characterization of one CrSBr sample of thickness \SI{19}{\nano\meter}, that we compare, in the following, with the absorption resulting from the imaginary part of the dielectric constant obtained from ab-initio calculations based on the Bethe-Salpeter Equation (BSE), reported in Fig.~\ref{fig3}b-c (Supplementary Information S3).

The Two-Photon PL (TP-PL) performed at RT has a spectral shape, characterized by a broad (\SI{97}{\milli\electronvolt} full-width at half maximum, FWHM) emission peak centered at $\approx$ \SI{1.31}{\electronvolt} (X, left panel of Fig.~\ref{fig2}a). When the sample is cooled down to LT (\SI{40}{\kelvin}), the TP-PL splits into sharp and narrow peaks with significantly decreased intensity (Supplementary Information S4), in agreement with previous PL measurements reported in literature~\cite{wilson2021interlayer}. The integrated TP-PL intensity at different excitation photon energies (right panel of Fig.~\ref{fig2}a) is significantly higher at RT compared to LT. At both temperatures, we observe a TP-PL resonance Y at \SI{1.82}{\electronvolt}.

Fig.~\ref{fig2}b shows the SH intensity from our sample at LT and RT for different excitation photon energies in the range 1.4$-$\SI{1.9}{\electronvolt} (Supplementary Information S5). Here, we observe a strong increase in the SH intensity for all photon energies, up to a factor of 25 at \SI{1.80}{\electronvolt}, below the N\'eel temperature ($\approx$\SI{132}{\kelvin}). This behavior is due to the phase transition from the centrosymmetric PM phase, to the non-centrosymmetric A-type AFM phase. Note that in the latter, both space inversion (IS) and time reversal symmetry (TRS) are broken \cite{lee2021magnetic}. The finite nonzero SH signal observed at RT (where IS is preserved in the PM phase) has been previously ascribed to the electric quadrupole response \cite{liu2022three}. 

In contrast to SHG, THG does not require broken IS, and it can thus also occur in the centrosymmetric PM phase of CrSBr. Indeed, we measured close to identical TH intensities at LT and RT for almost all photon energies, with the only exception of a clear temperature dependence for THG at $\approx$\SI{1.3}{\electronvolt}, \textit{i.e.} in correspondence with the lowest-lying optical transition (Fig.~\ref{fig3}b).

In SHG, we observe two resonances at approximately \SI{1.56}{\electronvolt} and \SI{1.8}{\electronvolt}, where the latter is close to the TP-PL exciton resonance. In THG, we observe a resonance X at \SI{1.3}{\electronvolt}, consistent with the PL emission. Moreover, THG reveals additional individual resonances at energies of \SIlist{1.96;2.07}{\electronvolt}. 

To gain more insights into the resonances observed in the SHG and THG measurements (Fig.~\ref{fig2}), we calculated the absorption of bulk CrSBr from $GW$/BSE calculations (Fig.~\ref{fig3}b-c). It is worth highlighting that NLO processes, such as SHG and THG, are enhanced in conjunction with real state (excitonic and/or electronic) transitions \cite{dogadov2022parametric}. Thus, the calculated absorption from BSE can be compared to the peaks observed in the SH and TH intensities at different excitation photon energies (Fig.~\ref{fig2}b-c).

By comparing the BSE results with the NLO experiments, we assign the lowest resonance X around \SI{1.31}{\electronvolt} to the lowest lying visible exciton (peak X$_\mathrm{A}$) in Fig.~\ref{fig3}b. We note that the conduction band splitting (CBM and CBM+1 in Fig.~\ref{fig3}a) results in two peaks X$_\mathrm{D}$ and X$_\mathrm{A}$, where X$_\mathrm{D}$ is the transition from the VBM into the CBM and X$_\mathrm{A}$ is the transition from the VBM to CBM+1. X$_\mathrm{D}$ is dark (i.e., forbidden in the dipole approximation) due to its orbital character. However, X$_\mathrm{A}$ and X$_\mathrm{D}$ slightly mix, leading to a small absorption also for X$_\mathrm{D}$, approximately three orders of magnitude smaller compared to the absorption of the bright transition X$_\mathrm{A}$.
The BSE calculations also reveal multiple excitonic features above \SI{1.7}{\electronvolt} along both the $\hat{\mathrm{a}}$ as well the $\hat{\mathrm{b}}$-axis. Since the excitation polarization is not aligned with either of the crystal axes, the resonance Y seen in both TP-PL and SHG measurements at \SI{1.8}{\electronvolt} (Fig.~\ref{fig2}a,b) is a mixture of features from $\hat{\mathrm{a}}$ and $\hat{\mathrm{b}}$ direction.

\subsection{\label{sec:3}Temperature dependence of the PL and NLO response}
We now focus on the temperature dependence of the measurements reported so far, namely PL, SHG and THG. 

\begin{figure}[htb!]
\includegraphics[width=\textwidth]{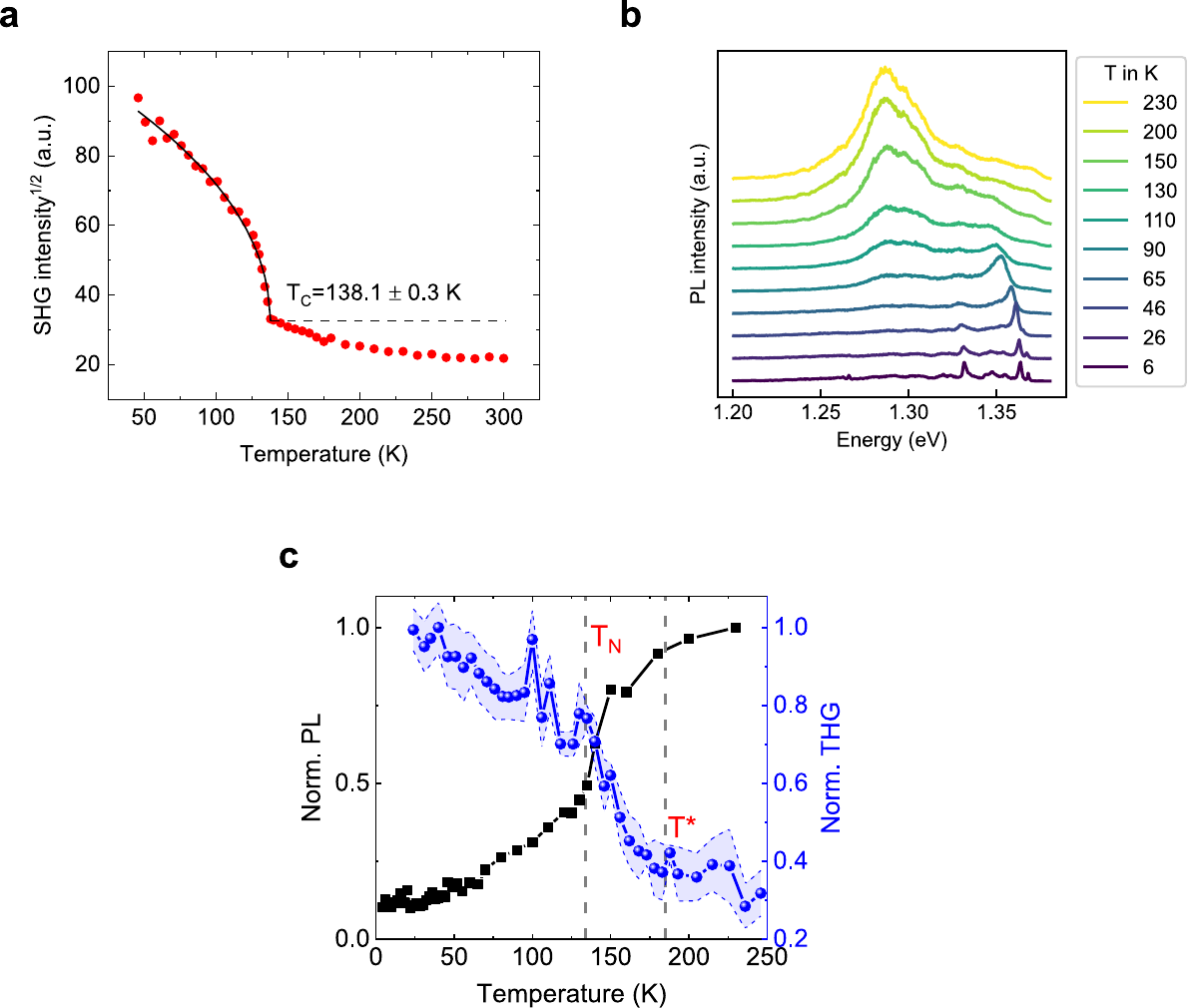}
\caption{\label{fig4}\textbf{Temperature dependence of NLO transitions and PL}. \textbf{a}, Temperature dependent square root of the SHG intensity measured at \SI{1.78}{\electronvolt}. The black solid line is a fit using the $(1-T/T_{\rm{C}})^{\beta}$ model of Eq.\ref{eqn:SHG_T_dep}. \textbf{b}, Temperature dependence of PL spectra showing strong intensity decrease below $T_{\rm{N}}$ and narrow peaks at low temperature. \textbf{c}, Temperature dependence of the normalized integrated PL spectra from \SI{6}{\kelvin} to \SI{230}{\kelvin} and normalized THG intensity measured at \SI{1.29}{\electronvolt}. The blue shaded area indicates the standard deviation from the fitting of the THG spectra. $T_{\rm{N}}$ and $T^*$ are \SIlist{132;185}{\kelvin}, respectively.}
\end{figure}

As already discussed, in layered magnets the SH temperature dependence is a sensitive probe of phase transitions, that manifest in breaking of both IS and TRS, and thus to the appearance of time-noninvariant terms in the nonlinear optical susceptibility \cite{lee2021magnetic}. In our CrSBr sample, we observe a clear temperature dependence of the SHG, as shown for instance in Fig.~\ref{fig4}a for a SH photon energy of \SI{1.78}{\electronvolt} (fundamental photon energy of \SI{0.89}{\electronvolt}), where the SH intensity decreases for increasing temperature. When bulk CrSBr undergoes a phase transition from PM to AFM, its SH intensity can be modeled as~\cite{lee2021magnetic}: 

\begin{align}
\label{eqn:SHG_T_dep}
    I_{\rm SHG}^{1/2} \propto (1-T/T_{\rm{C}})^\beta
\end{align}

where $T_{\rm{C}}$ is the critical temperature and $\beta$ the critical exponent. In Fig.~\ref{fig4}a we plot the square root of the total SH intensity as a function of temperature and fit it with Eq.\ref{eqn:SHG_T_dep}, where we use $T_{\rm{C}}$ and $\beta$ as fitting parameters. From this, we obtain a critical temperature of \SI{138.1 \pm 0.3}{\kelvin} and $\beta = 0.41\pm0.05$, in good agreement with previous theoretical and experimental reports~\cite{goser1990magnetic,wang2019family,telford2020layered,torres2023probing}. Note that the slight temperature dependence of SHG above $T_{\rm{C}}$ is not fully understood yet. As this is not the main focus of our current paper, we postpone a more detailed investigation to future work.

The temperature dependence of the PL and TH intensities (measured at \SI{1.29}{\electronvolt}) follow a different trend. Both display a strong temperature dependence, with a step variation in correspondence with the N\'eel temperature, although in opposite directions (Fig.~\ref{fig4}c). The integrated PL intensity increases with increasing temperature, while the TH intensity decreases for increasing temperature, following two distinct transitions at temperatures of $T_{\rm{N}}=$\SI{132}{\kelvin} and approximately at the previously reported temperature of the decay of intra-layer long range ordering $T^*=$\SI{185}{\kelvin} \cite{lin2024strong}. Recent measurements on bulk CrSBr have described the magnetic phase transition as a three-stage process: (1) above $T^*$ the spin orientations of PM order are randomly distributed; (2) when cooling down, long-range magnetic orders are established, and the magnetic property is characterized by an intermediate ferromagnetic phase (iFM), where in-plane FM order is established while the magnetic order of interlayers is incommensurate~\cite{liu2022three}; (3) for temperatures below $T_{\rm{N}}$ the layers couple antiferromagnetically. Thus, during the heating process from \SI{24}{\kelvin} to RT, the CrSBr sample will undergo a three-stage phase change, from the AFM phase below $T_{\rm{N}}$, to iFM phase between $T_{\rm{N}}$ and $T^*$ and finally to the random PM order above $T^*$~\cite{liu2022three}. Looking at Fig.~\ref{fig4}c, these different phase transitions, which ultimately connect to the spin orientation in the CrSBr sample, have an effect on the resonant TH intensity.

Finally, we discuss the opposite temperature dependence of PL and TH intensities. First, we note that the TH resonant enhancement directly follows the linear absorption of the sample, while, in contrast, the PL intensity is due to the combined effect of absorption efficiency (at the excitation photon energy) and radiative emission rate. Thus, an enhanced TH intensity below $T_{\rm{N}}$ indicates that the absorption at \SI{1.29}{\electronvolt} is higher in the AFM compared to the PM phase. In parallel, the reduced PL emission below $T_{\rm{N}}$ must indicate the presence of additional and efficient non-radiative decay channels in the AFM configuration compared to the PM phase. 

Non-radiative decay channels in a highly ordered magnetic phase could hypothetically either be assigned to a spin mediated/enhanced decay, for instance due to strong spin-phonon coupling effects, which certainly play a role in CrSBr~\cite{torres2023probing,pawbake2023raman}, or to the appearance of low-energy dark states, which could trap excitons and thus quench the PL intensity at low temperatures~\cite{klein2023bulk,Tabataba-Vakili2023}, similar to what has been observed \textit{e.g.} in monolayer WSe$_{\rm{2}}$~\cite{zhang2015}. However, while in TMDs the dark state is a spin-forbidden transition, in CrSBr the lowest energy dark state responsible for the PL quenching could be the X$_\mathrm{D}$ transition obtained from BSE calculations and reported in Fig.~\ref{fig3}b. Recent measurements have reported the appearance of a faster exciton PL decay time in CrSBr for decreasing temperatures~\cite{lin2024strong}. Combined with the observation of an overall reduction of the PL yield at low temperatures, the faster decay dynamics can only be attributed to non-radiative recombination pathways, as already mentioned above. Our results fully support this scenario: when tuning the temperature from RT to LT, the TH intensity increases due to an increase of the oscillator strength/absorption~\cite{Shao2025}, while the PL intensity decreases due to the emergence of non-radiative decay channels~\cite{lin2024strong}. 

\subsection{Conclusion}
In conclusion, we have investigated the wavelength- and temperature-dependent linear and nonlinear optical properties of layered CrSBr across its paramagnetic and antiferromagnetic phases. Nonlinear spectroscopy revealed clear resonances in both second- and third-harmonic generation, which we compared with ab initio calculations to assign direct-gap optical transitions. We further examined the temperature dependence of second-harmonic generation, third-harmonic generation, and linear photoluminescence. The second-harmonic response shows a pronounced temperature dependence across the entire probed spectral range, consistent with the breaking of inversion symmetry below the N\'eel temperature. In contrast, the third-harmonic intensity is largely insensitive to temperature, except at resonance with the lowest bright optical transition. By comparing the temperature dependence of resonant third-harmonic generation and linear photoluminescence at this transition, we infer the onset of efficient non-radiative recombination in the antiferromagnetic phase, possibly related to strong spin-phonon coupling or to the ultrafast decay to a low energy optically dark state.

\subsection{Acknowledgments}

T.D. acknowledges financial support from the Deutsche Forschungsgemeinschaft (DFG, German Research Foundation) through Project No. 426726249 (DE 2749/2-1 and DE 2749/2-2). The authors gratefully acknowledge the Gauss Centre for Supercomputing e.V. (www.gauss-centre.eu) for funding this project by providing computing time through the John von Neumann Institute for Computing (NIC) on the GCS Supercomputer JUWELS~\cite{Alvarez2021} at Jülich Supercomputing Centre (JSC).

Z.S. was supported by ERC-CZ program (project LL2101) from Ministry of Education Youth and Sports (MEYS) and by the Advanced Multiscale Materials for Key Enabling Technologies project, supported by the Ministry of Education, Youth, and Sports of the Czech Republic. Project No. CZ.02.01.01/00/22\_008/0004558, Co-funded by the European Union.

B.P. acknowledges financial support from the US Department of Energy, Office of Science, Basic Energy Sciences, Materials Sciences and Engineering Division through Argonne National Laboratory under contract no. DE-AC02-06 CH11357.

J.K. acknowledges funding through the National Science Foundation (NSF) under Trailblazer Engineering Impact Award No. 2421694 and support from the U.S. Department of Energy, Office of Science, Office of Basic Energy Sciences, Division of Materials Sciences and Engineering under Award No. DE-SC0025387.

G.S. acknowledges funding from the Deutsche Forschungsgemeinschaft (DFG, German Research Foundation) via the SFB 1375 NOA “Nonlinear Optics down to Atomic scales” (project number 398816777), IRTG 2675 ”META-ACTIVE” (project number 437527638 and project number 448835038) and WHAT-A-TWIST (project number 547611111).

\clearpage
\bibliography{references.bib}

\end{document}


\section{Title}
\noindent
Supplementary Information - Probing temperature dependent non-radiative decay channels in CrSBr  with nonlinear optics

\section{Author list}
\noindent
Minjiang Dan$^{1,2,3,10}$, Till Weickhardt$^{2,10}$, Paul Herrmann$^2$,  Fabian Glatz$^2$,  Marie-Christin Heißenbüttel$^4$, Thorsten Deilmann$^4$, Michael Rohlfing$^4$, Ksenia Mosina$^5$, Zden\v ek Sofer$^5$, Benjamin Pingault$^{6,7}$, Julian Klein$^{8,\star}$, Giancarlo Soavi$^{2,9,\dagger}$

\section{Affiliations}
\noindent
$^1$Joint Laboratory for Extreme Conditions Matter Properties, School of Mathematics and Physics, Southwest University of Science and Technology, Mianyang 621010, China
\newline
$^2$Institute of Solid State Physics, Friedrich Schiller University Jena, Helmholtzweg 5, 07743 Jena, Germany
\newline
$^3$School of Physics, University of Electronic Science and Technology of China, Chengdu 610054, China
\newline
$^4$Institute of solid state theory, University of Münster, 48149 Münster, Germany
\newline
$^5$Department of Inorganic Chemistry, University of Chemistry and Technology Prague, Technickà 5, 166 28, Prague 6, Czech Republic
\newline
$^6$Materials Science Division and Q-NEXT, Argonne National Laboratory, Lemont, IL, USA
\newline
$^7$Chicago Quantum Institute and Pritzker School of Molecular Engineering, University of Chicago, Chicago, IL, USA
\newline
$^8$Department of Materials Science and Engineering, Massachusetts Institute of Technology, Cambridge, Massachusetts 02139, USA
\newline
$^{9}$Abbe Center of Photonics, Friedrich Schiller University Jena, Albert-Einstein-Straße 6, 07745 Jena, Germany
\newline
$^{10}$These authors contributed equally to this work. 
\newline
$^{\star}$ jpklein@mit.edu
$^{\dagger}$ giancarlo.soavi@uni-jena.de

\section{S1 Synthesis of bulk crystals}
Chromium (99.99\%, -60 mesh, Chemsavers, USA), sulfur (99.9999\%, Stanford Materials, USA) and bromine (99.9999\%, Sigma-Aldrich, Czech Republic) were mixed referring to the stochiometric ratio in a quartz ampoule and then reacted directly, in order to synthesize bulk CrSBr crystals. While keeping the second end of the ampoule below \SI{250}{\celsius}, the compound was pre-reacted for 12 hours at \SI{700}{\celsius} inside a crucible furnace. This procedure is redone to the absence of liquid bromine. Subsequently, the ampoule is placed in a two-zone horizontal furnace. In the first step the growth zone was heated to \SI{900}{\celsius} and the source zone to \SI{700}{\celsius} for 25 hours. In the second step, the crystals were grown for 7 days by heating the source zone to \SI{940}{\celsius} and the growth zone to \SI{850}{\celsius}. Finally, the crystal was taken out in an Ar glovebox~\cite{Klein2022}.

\section{S2 Sample fabrication and characterization}
\begin{figure}
\includegraphics[width=\textwidth]{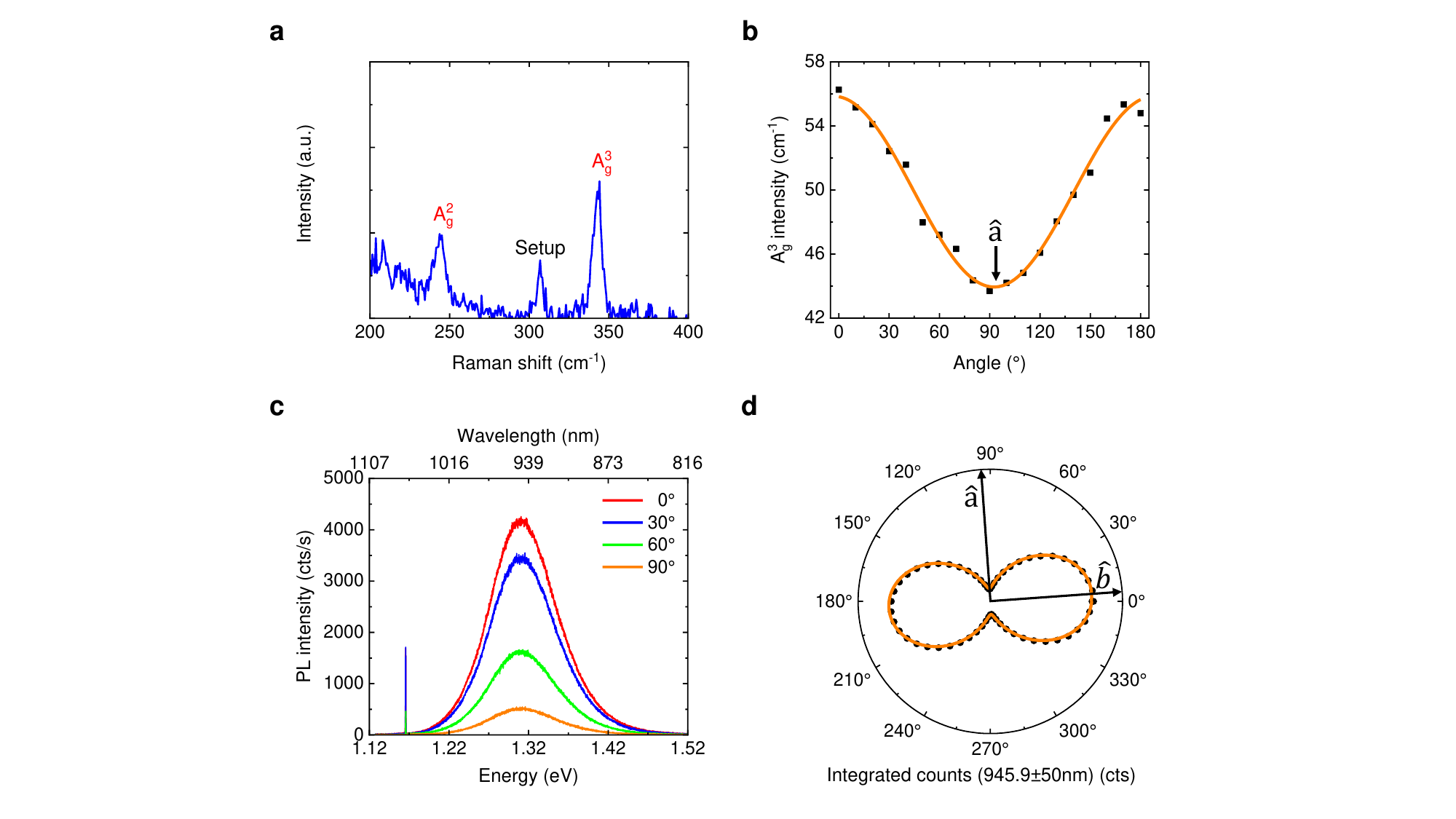}
\caption{\label{fig S1} \textbf{Raman and PL characterization at room temperature}. \textbf{a}, Raman spectrum from \SI{200}{\centi\meter^{-1}} to \SI{400}{\centi\meter^{-1}}. \textbf{b}, Polarization dependence of the A$_{\rm{g}}^3$ mode of Raman mode. \textbf{c}, PL spectra under different polarization angles excited by a \SI{2.33}{\electronvolt} laser. \textbf{d}, Polarization dependence of the PL intensity.}
\end{figure}

CrSBr samples were further fabricated through mechanical exfoliation and then transferred onto the SiO$_{\rm{2}}$(\SI{90}{\nano\meter})/Si substrate. For sample characterization, Raman and PL measurements were performed by exciting the sample with a \SI{2.33}{\electronvolt} CW laser (Cobolt 08-DPL, Hübner photonics) and detected by a spectrometer (iHR550, Horiba). Due to the broad notch filter (NF533-17, Thorlabs), we only observe the A$_g^2$ mode $\sim$ \SI{244.2}{\centi\per\meter} and  A$_g^3$ mode $\approx$ \SI{ 343.8}{\centi\per\meter}, while we ascribe the line at $\approx$ \SI{307}{\centi\per\meter} to an artifact from the setup (Fig.~S\ref{fig S1}a). We rotate the polarization axis of the excitation laser with a half-wave plate (WPH10M-532, Thorlabs) and detect the intensity of the A$_g^3$ mode (Fig.~S\ref{fig S1}b) as well as PL (Fig.~S\ref{fig S1}d). Both polarization dependencies are consistent with each other, showing the same minimum/maximum emission angles. This is in agreement with the quasi 1-dimensional nature of CrSBr, and from this we can infer that the crystal $\hat{\mathrm{b}}$-axis is slightly tilted with respect to the laboratory frame, see Fig.~S\ref{fig S1}b-d. 

\subsection*{S3 Electronic structure and optical excitations from first principles}
Our theoretical approach to investigate CrSBr has been discussed in Ref.~\cite{klein2023bulk}.
In the following, we briefly recapitulate the main ingredients and the numerical evaluation.
The starting point for our calculations is the density functional theory (DFT) within the generalized gradient approximation (GGA) \cite{gga_pbe}.
Because CrSBr exhibits spin–orbit coupling and in-plane ferromagnetism, we employ the noncollinear version of DFT \cite{Sandratskii_1998,so_Staerk_2011} to account for the magnetic structure and spin–orbit coupling, yielding spinor wave functions as a starting point for the subsequent many-body calculations.

Quasiparticle (QP) band structures were obtained from many-body perturbation theory within the $GW$ approximation \cite{HedinGW}.
The electronic self-energy includes the one-particle Green's function and the screened Coulomb interaction, in which the dielectric screening of the latter is obtained within the random-phase approximation.
For CrSBr, the off-diagonal contributions to the self-energy are important; therefore, the QP wave functions were determined by diagonalizing the self-energy in the basis of DFT wave functions \cite{Foerster_2015,O2MoS2}.
To accelerate the self-consistent determination of the QP corrections, a scissor operator of 1.0\,eV was used to anticipate the opening of the band gap of bulk CrSBr.
To evaluate the optical behavior including the electron-hole interaction, we solve the Bethe-Salpeter equation (BSE) \cite{Rohlfing_eh}.

In the $GW$ calculations, the two-point functions were represented using a hybrid basis of Gaussian orbitals and plane waves, with a plane-wave cutoff of 1.5\,Ry.
The $k$-mesh is chosen to be $12\times 9 \times 4$ for the random phase approximation, $30\times 22\times 4$ for the evaluation of the $GW$ self-energy, and a mesh of $16 \times 12 \times 4$ is used for the BSE.
For the optical spectra an artificial broadening of 35\,meV is applied.

\section{S4 Temperature dependent photoluminescence spectroscopy}
We use a helium closed-cycle cryostats (Montana Instruments or AttoDry 800) for the temperature dependence of the PL. The sample is accessed through a side window in both setups, where we use a home-built confocal microscope with a ×100, 0.9 NA objective (Olympus) to focus the excitation laser and collect the PL after using a long-pass filter. The emitted light is coupled into a fibre and spectrally resolved in a high-resolution spectrometer, where it is detected by a liquid nitrogen cooled charge-coupled device camera.

\section{S5 Nonlinear optics}

\subsection{S5.1 Experimental setup}
For the wavelength-dependent measurements, we take the signal (from \SI{0.62}{\electronvolt} to \SI{0.95}{\electronvolt}) and idler (from \SI{0.31}{\electronvolt} to \SI{0.59}{\electronvolt}) from an optical parametric oscillator (Levante IR fs, APE) pumped by a high-power Yb-doped femtosecond oscillator (FLINT FL2-12, Light Conversion) with \SI{76}{\mega\hertz} repetition rate and $\sim$\SI{100}{\femto\second} pulse duration. During the whole measurement, we put the sample inside a liquid helium-flow cryostat (ST-500, Janis) with coupled silicon temperature controller (Model 325, Lake Shore) to monitor and control the sample temperature. The laser is focused onto the sample through a reflective objective (LMM-40X-P01, Thorlabs, 40$\times$, 0.50 NA) and then the backward scattered SHG and THG signals are separated by a dichroic mirror and shortpass filter (DMSP950, Thorlabs; FESH0950, Thorlabs) while a beamsplitter (BSW23, Thorlabs) is used for TP-PL. Finally, SHG, THG and TP-PL are detected by a spectrometer (iHR320, Horiba). The power dependence of these NLO processes is measured using a single-photon-avalanche-diode (SPAD, C11202-050, Hamamatsu) with appropriate bandpass filters.

\subsection{S5.2 Excitation power dependence}
\begin{figure}
\includegraphics[width=\textwidth]{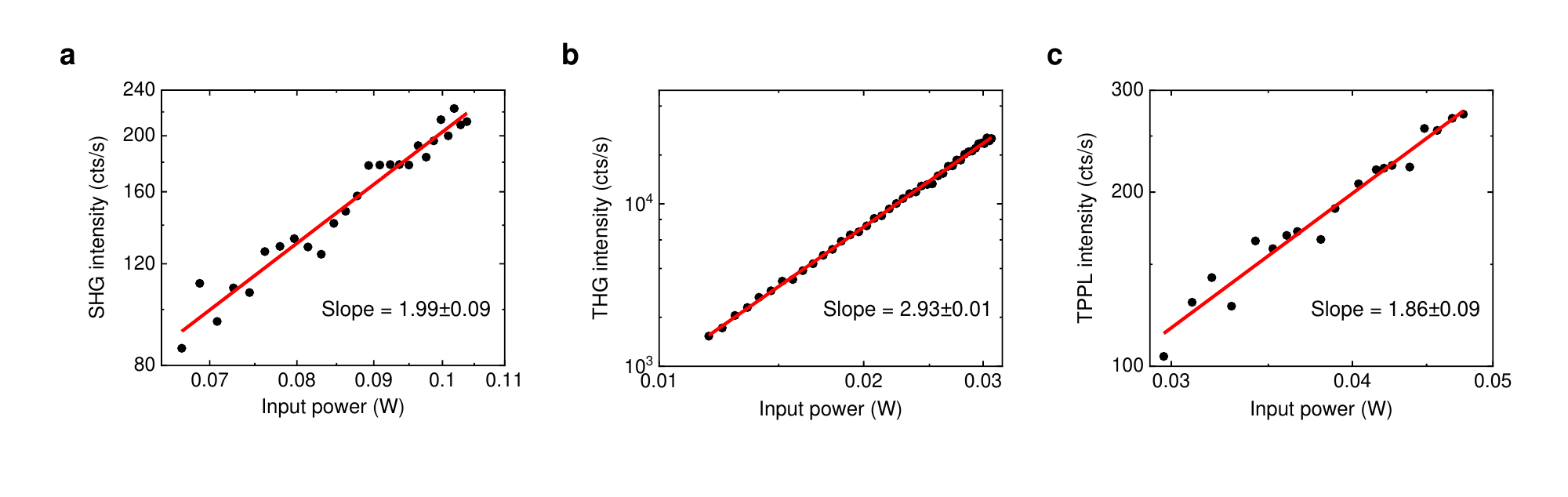}
\caption{\label{fig S2} \textbf{Excitation power dependence of the NLO processes}. Excitation power dependence of SHG (\textbf{a}), THG (\textbf{b}) and TP-PL (\textbf{c}) intensities.}
\end{figure}

Fig.~S\ref{fig S2} shows the excitation power dependence of the different NLO processes. In order to maximize the signal, SHG and THG are measured at \SI{17}{\kelvin} with incident photon energies of \SI{1.55}{\electronvolt} and \SI{2.41}{\electronvolt}, respectively,  while TP-PL is measured at room temperature with an incident photon energy of \SI{1.82}{\electronvolt}. Upon fitting the experimental data with a power law $I(SHG/THG/TP-PL)\propto I(FF)^n$, where the characteristic exponent $n$ is the fitting parameter, we retrieve the expected quadratic dependence for both  SHG (Fig.~S\ref{fig S2}a) and TP-PL (Fig.~S\ref{fig S2}c), and cubic dependence for THG (Fig.~S\ref{fig S2}b). Small deviations from the expected power dependence could be ascribed to ultrafast coherent bandgap modulation induced by the fundamental beam~\cite{klimmer2026}.

\subsection{S5.3 Wavelength dependence}
\begin{figure}
\includegraphics[width=\textwidth]{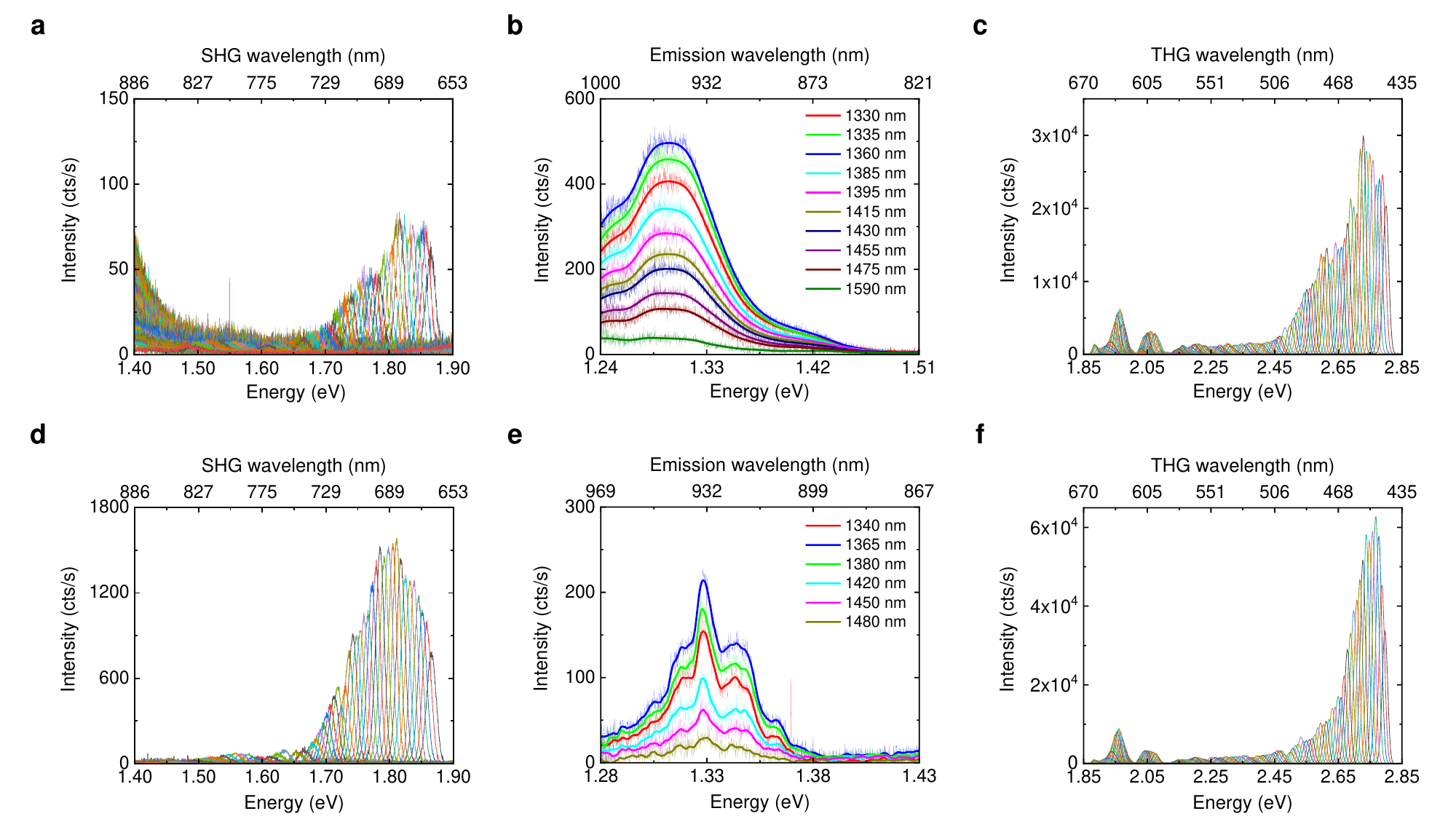}
\caption{\label{fig S3} \textbf{NLO spectra from OPO signal}. Room temperature SHG spectra (\textbf{a}),  TP-PL spectra (\textbf{b}), THG spectra (\textbf{c}). Low temperature (\SI{40}{\kelvin}) SHG spectra (\textbf{d}), TP-PL spectra (\textbf{e}) and THG spectra (\textbf{f}).}
\end{figure}

Fig.~S\ref{fig S3} shows all NLO spectra obtained upon excitation with the signal output of the OPO (see section S4.2) at room temperature (Fig.~S\ref{fig S3}a-c) and low temperature (Fig.~S\ref{fig S3}d-f). We sweep the laser wavelength in \SI{5}{\nano\meter} steps and keep the power on the sample fixed. Since the efficiency of SHG (Fig.~S\ref{fig S3}a,d) and TP-PL (Fig.~S\ref{fig S3}b,e) are low compared to THG (Fig.~S\ref{fig S3}c,f), we excite SHG and TP-PL with an average power of \SI{50}{\milli\watt}, compared to \SI{15}{\milli\watt} for THG.

\begin{figure}
\includegraphics[width=\textwidth]{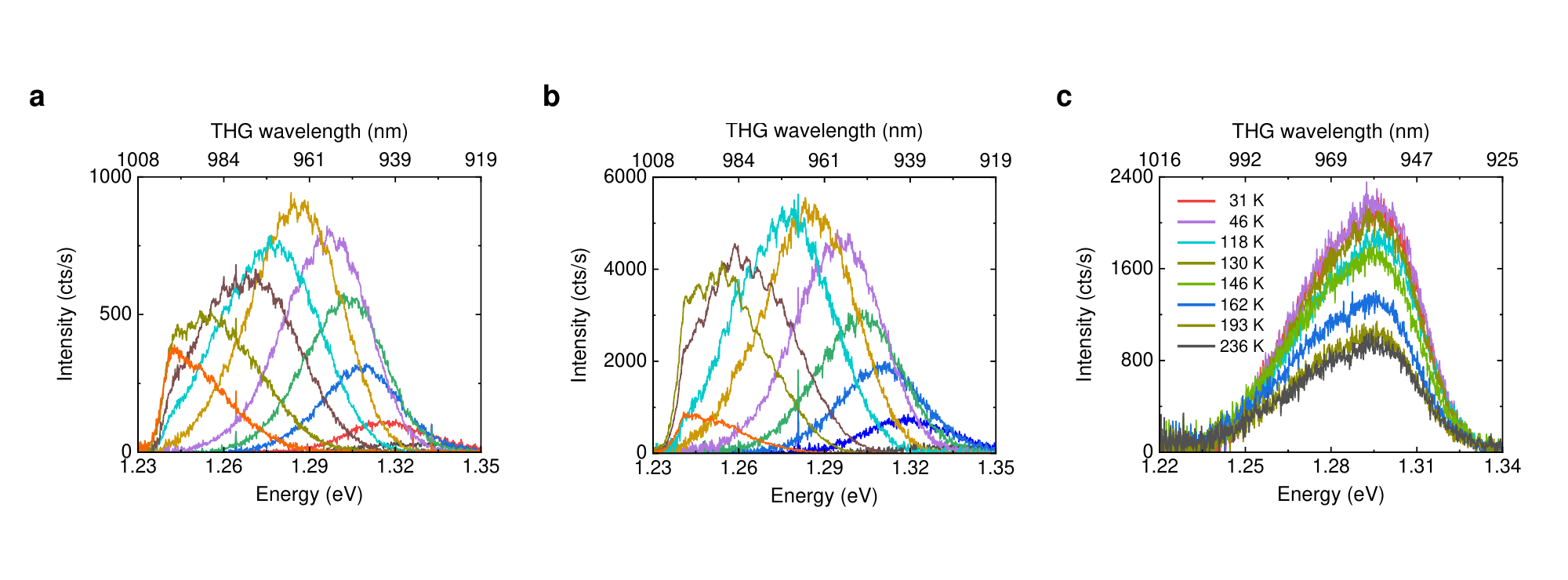}
\caption{\label{fig S4} \textbf{Temperature dependent THG spectra from OPO idler}. THG spectra for a fundamental photon energy from \SI{0.413}{\electronvolt} to \SI{0.454}{\electronvolt} at RT (\textbf{a}) and low temperature (\textbf{b}). \textbf{c}, THG spectra for emission at photon energy of \SI{1.29}{\electronvolt} as a function of temperature.}
\end{figure}

In order to characterize the region of the PL peak, we also couple the OPO idler output into our setup and measure THG spectra as a function of the laser excitation energy. We get a good agreement between PL and THG, as shown in Fig1.d,f of the main text. The TH intensity increases at low temperature compared to RT as shown in Fig.~S\ref{fig S4}a,b and further discussed in the main text. The spectra are collected on a spectrometer after a shortpass filter (FESH 1000, Thorlab).

\bibliography{references.bib}